# Leveraging Machine Learning for Industrial Wireless Communications

Ilaria Malanchini, Patrick Agostini, Khurshid Alam, Michael Baumgart, Martin Kasparick, Qi Liao, Fabian Lipp, Nikolaj Marchenko, Nicola Michailow, Rastin Pries, Hans Schotten, Slawomir Stanczak, Stanislaw Strzyz

*Abstract*—Two main trends characterize today's communication landscape and are finding their way into industrial facilities: the rollout of 5G with its distinct support for vertical industries and the increasing success of machine learning (ML). The combination of those two technologies open the doors to many exciting industrial applications and its impact is expected to rapidly increase in the coming years, given the abundant data growth and the availability of powerful edge computers in production facilities. Unlike most previous work that has considered the application of 5G and ML in industrial environment separately, this paper highlights the potential and synergies that result from combining them. The overall vision presented here generates from the KICK project, a collaboration of several partners from the manufacturing and communication industry as well as research institutes. This unprecedented blend of 5G and ML expertise creates a unique perspective on ML-supported industrial communications and their role in facilitating industrial automation. The paper identifies key open industrial challenges that are grouped into four use cases: wireless connectivity and edge-cloud integration, flexibility in network reconfiguration, dynamicity of heterogeneous network services, and mobility of robots and vehicles. Moreover, the paper provides insights into the advantages of ML-based industrial communications and discusses current challenges of data acquisition in real systems.

## I. INTRODUCTION

One of the most important differences between 5G and previous generations of cellular networks lies in the strong focus of 5G on supporting vertical industries. Connected driving, agriculture, logistics, and health are undoubtedly important target businesses, but today, industrial automation, production facilities, and the industrial Internet of Things are seen as the most demanding and, probably, most attractive new application domains [1]. The capabilities of 5G extend far beyond mobile broadband with ever increasing data rates. Specifically, 5G supports communications with unprecedented reliability and very low latency, as well as massive IoT connectivity. This paves the way for numerous exciting industrial applications including new levels of human-robot interaction, advanced concepts of human-machine collaboration, support for fleets of automated guided vehicles (AGVs), and collaborative robots. As enabler for these new applications and as the basis for a new generation of holistic industrial connectivity solutions, 5G might even have a disruptive impact when related building blocks, such as wireless connectivity, edge computing, or network slicing, are introduced into future smart factories.

With artificial intelligence and, in particular, machine learning (ML), another technology besides 5G, with the potential to dramatically change the way we design and operate networks, is now finding its way into industrial facilities [2]. Depending on the application, ML has in the last years proven to be a powerful tool for classification of events, detection of anomalies, prediction, and tactical decision making.

This work was supported by the German Federal Ministry of Education and Research (BMBF) project KICK.
Corresponding author: Ilaria Malanchini, e-mail: ilaria.malanchini@nokia-bell-labs.com.
Ilaria Malanchini, Qi Liao, and Rastin Pries are with Nokia; Patrick Agostini, Martin Kasparick, and Slawomir Stanczak are with Fraunhofer Heinrich Hertz Institute and Technische Universität Berlin; Khurshid Alam and Hans Schotten are with the German Research Center for Artificial Intelligence; Hans Schotten is also with the University of Kaiserslautern; Michael Baumgart and Fabian Lipp are with Infosim; Nikolaj Marchenko is with Robert Bosch GmbH; Nicola Michailow is with Siemens Corporate Technology; Stanislaw Strzyz is with atesio.

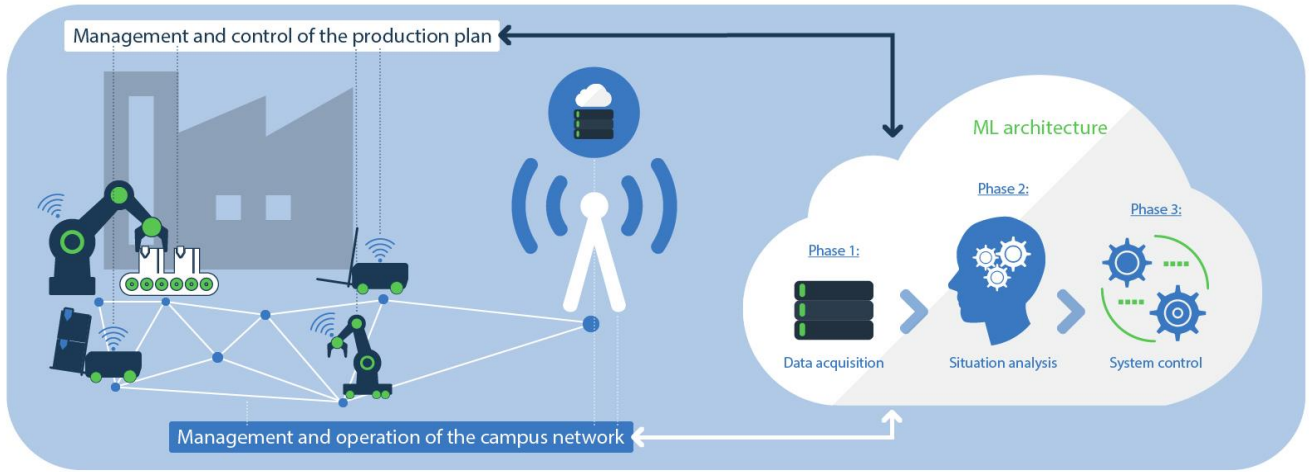

Fig. 1. ML enables synergies of production and communication worlds in industrial environments.

Unlike most of the existing work, which either addresses the challenges of applying 5G technologies to industrial communication systems [3], or provides a review on data mining and ML approaches in the manufacturing industry [4], the purpose of this paper is to highlight the potential and synergies resulting from the combination of these two major technologies (i.e., 5G and ML) in industrial environments. In the following, the overall vision developed and promoted within the context of the project "KICK" [5] is presented. The project consortium brings together diverse partners, from manufacturing and communication industries to research institutes with a strong ML expertise. This creates a unique perspective on ML-empowered industrial communications, hereafter also referred to as *campus networks*, and their role in enabling industrial automation. Since network traffic and processes in the physical world, particularly in controlled environments such as production facilities, are highly correlated, understanding, classifying, and predicting their behaviors will introduce a joint intelligence from which information technology (IT) and operational technology (OT) will highly benefit. Fig. 1 illustrates the combination of the production and communication worlds. Data are collected from both the production and network management systems, they are analyzed and used to optimize and adapt the campus network to production needs.

The remainder of the paper elaborates on the KICK vision and presents what the consortium believes are the open industrial challenges by classifying them into four key use cases:

- Wireless connectivity and edge cloud integration
- Flexibility in campus network reconfiguration
- Dynamicity of heterogeneous network services
- Mobility of robots and vehicles

The paper also provides insights on the benefits of ML-based campus networks and discusses current challenges of data acquisition in real systems. The contribution is based on two guiding questions: (i) What are the benefits of 5G and ML in terms of support to factory modernization, increased flexibility, improved efficiency and better performance for the IT and OT side? (ii) Which functionalities are needed from communication and ML perspectives to support such industrial challenges?

## II. INDUSTRIAL USE CASES

Manufacturing is experiencing a trend from mass-production to mass-customization [6], with time and cost of reconfiguring production lines posing major challenges. To achieve profitable operation with a batch-size of one, factory owners are investing heavily in digitalization, machines are becoming more versatile, human-machine collaboration is becoming more sophisticated, and fixed production lines are replaced by flexible production stations.

In the following, specific use cases are derived from selected industrial challenges that are, on the one hand, highly important for the production industry and, on the other hand, appear promising to be solved by the adoption of 5G and the application of ML techniques. Each use case describes the real challenges or desired development directions of the modern production enterprises. These are then translated into envisaged network solutions, supported by suitable ML techniques. Fig. 2 shows the transition from traditional, mainly wired static factories to future wirelessly interconnected flexible factories, enabled by the presented use cases. Table I gives a concise overview of those, highlighting the key elements: industrial challenge, communication solution, and ML support.

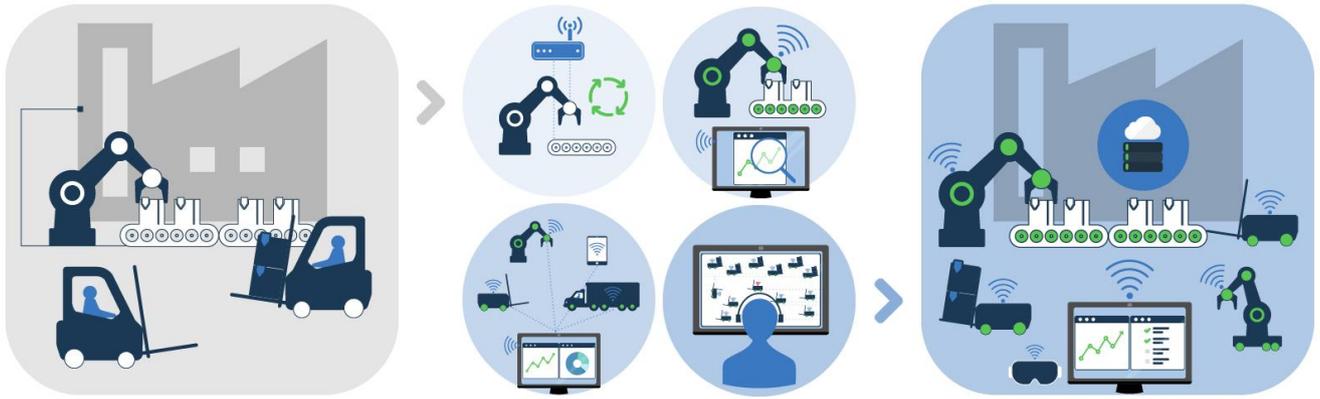

Fig. 2. Transition from traditional to future factory enabled by the four use cases (from top left and clockwise: wireless connectivity and edge cloud integration, flexibility in campus network reconfiguration, mobility of robots and vehicles, and dynamicity of heterogeneous network services).

## A. Wireless Connectivity and Edge Cloud Integration

Industrial control applications constitute one of the central and critical parts of the factory automation. In simple terms, a control application consists of a controller, a controlled system (actuators), and one or multiple sensors. Control applications are typically deployed on dedicated real-time programmable logic controllers (PLCs) or industrial PCs, which are often placed next to the sensors and actuators. When hundreds of such controllers are in use on a factory floor, it is challenging to manage, upgrade, and scale such systems cost-effectively. Furthermore, with the increasingly required flexibility, re-cabling production machines and reconfiguring networks and applications become complex, time-consuming and expensive tasks.

Introducing wireless connectivity increases the *flexibility* of the factory floor and the *scalability* of control systems. Furthermore, the high reliability and low latency promised by 5G systems guarantee network performance suitable for many control applications, such as mobile robotics. Another promising solution is the use of an on-premise edge cloud with a highly virtualized infrastructure [7]. Applications like closed-loop control, simultaneous localization and mapping (SLAM), and rendering of augmented reality (AR) generate multimedia traffic with high demand on both network resources for transmission and computing resources for post-processing. Thus, they greatly benefit from the introduction of a factory edge cloud, thanks to the increased shared computing resources and an easier to upgrade and maintain infrastructure. Although the combination of wireless connectivity and on-premise edge computing enables a powerful and flexible operation of industrial systems, a significant level of orchestration is required across computing, networking, and applications.

The orchestration can be supported by anticipating changes and needs, either in the network or in the production environment, and by exploiting those to perform application-level adaptations. For instance, the following two cases can be considered: (i) Self-adaptation of the application to the known or predicted conditions of the campus network, e.g., considering quality of service (QoS), coverage, traffic load, or fading; (ii) Initial deployment or dynamic offloading of specific functions (e.g., control, AR rendering) to the edge cloud based on the performance required by the applications and current network conditions.

In general, given the complexity and dynamic nature of the factory environment, application requirements, and data traffic, automatization and optimization require the adoption of ML approaches. A challenging task before applying ML approaches is to find a proper classification of different system states and associated traffic patterns. Such classification can help to predict bottlenecks and avoid them by proper load balancing, i.e., by dynamically offloading computationally expensive tasks from the devices to the edge cloud. Having found such a classification scheme, ML approaches such as reinforcement learning (RL) can be applied for predicting load changes or for optimizing the function placement and scheduling.

TABLE I
SUMMARY OF THE INDUSTRIAL USE CASES

| Use Case | Industrial Challenge | Communication Solution | Machine Learning Support |
|---|---|---|---|
| A | Scalability of industrial control systems and flexibility of the factory floor | 5G wireless connectivity and edge computing | *Prediction methods* to allow proactive adaptation of industrial applications based on predicted network and IT conditions, and *classification methods* to improve dynamic offloading to the edge cloud based on classified system states and traffic patterns |
| B | Flexibility of production facilities and adaptability to changes with minimum downtimes | Automated network planning and reconfiguration based on radio and reliability maps | *Object recognition* and *SLAM* algorithms to enable 3D reconstruction of the industrial environment for network planning, and *transfer learning* for network reconfiguration |
| C | Dynamicity of heterogeneous services, while providing reliability and isolation | Automated creation and management of virtually isolated network slices | *RL algorithms* for learning optimal slice parameters configuration and resource allocation, and *unsupervised learning* for anomaly detection |
| D | Mobility and management of robots and vehicles, while guaranteeing safety and security | Proactive network management system based on accurate radio environment tracking and device localization | *Pattern recognition* algorithms and *correlation analysis* to learn the mobility patterns and support QoS prediction and resource management, and *distributed* or *federated learning* to balance computing and network resources |

## B. Flexibility in Campus Network Reconfiguration

The increasing dynamicity of future industrial production strategies that results from the need for accommodating highly optimized and individualized processes, requires flexible production facilities capable of adapting efficiently and quickly to changing requirements and configurations with minimum downtimes. The resulting tightened needs on the *flexibility* and *adaptability* of the production infrastructure are directly reflected in the communication networks. As for the previous use case, wireless connectivity is an essential component of campus networks, but this also implies an inherent sensitivity of the communication systems to the frequent variations of the radio channel conditions, which are a direct result of the changes in the production environment. This calls for highly customized network planning and configuration solutions that empower plant operators to efficiently plan and manage the campus network at a minimum level of expertise.

In contrast to classical cellular networks, where the focus is on coverage optimization, the challenges faced in industrial communications are driven by the stringent reliability and latency constraints and by the diverse QoS requirements of industrial processes e.g., AR applications. Meeting those requires configurations of the campus networks that are optimized for the specific radio conditions of the production environment. For planned changes, such as the redesign of a factory floor or replacement of production lines, the impact on network performance must be determined in advance. Subsequent network reconfigurations should happen fully automated without human intervention. However, it should be guaranteed that all operations can be supervised, and human interaction is always possible when needed. Furthermore, adaptations need to be robust and reliable to prevent overhead and network instability caused by frequent reconfigurations.

A promising approach to support flexible campus network reconfigurations is based on a virtual, simulated representation of the factory floor, also known as *Digital Twin*. In a first phase, this requires the reconstruction of the industrial environment, composed by different objects, such as walls or machines. Video and lidar based scanning of the environment in combination with ML methods, such as deep-learning based object recognition, 3D object reconstruction, and SLAM algorithms are very promising for this task [8]. In a second phase, the reconstructed environments are utilized as input for model-driven radio propagation simulators, e.g., ray tracing, to estimate corresponding 3D radio maps that describe the achievable spatial network performance. Furthermore, *reliability maps* can be generated by fusing 3D radio maps with the prediction of other relevant metrics, e.g., traffic load, that can be leveraged to ensure that network configurations will support the stringent reliability and latency requirements of mission-critical traffic.

In case of reconfigurations of the production environment, which are often characterized by spatio-temporal changes in the communication traffic, radio propagation properties, and intralogistics flows, previously trained ML models become outdated and, in the worst case, must be retrained using data gathered from the new radio environment. To reduce the cost of retraining, both network and production data can be used to learn the impact of specific network configuration parameters on the performance and can be transferred among different production setups to predict how some changes on such parameters will affect the performance when no or little data from the new setup is available. In this respect, transfer learning, which allows the use of previously trained models, can be employed to leverage known and optimized configurations (and/or model-based configurations) to speed up the optimization process and reduce complexity and the amount of required training data [9].

*C. Dynamicity of Heterogeneous Network Services*

An additional aspect that characterizes industrial facilities is the coexistence of several, heterogeneous services that need to be managed by the same network infrastructure. In fact, within a campus there might be (i) production machines which generate logs along with alarms and errors, (ii) moving vehicles which need to be instructed on where to go and what to do, (iii) video cameras that need to record and send videos in real time to the edge cloud, (iv) trucks coming from the outside for loading or unloading goods, and (v) workers with their own devices. Furthermore, the industrial environment is highly *dynamic* making all these classes of services varying in time and space. Additional challenges include sudden changes in the traffic load, for instance due to arrival of many devices or a shift change in the factory. When managing such heterogenous traffic, specific requirements need to be fulfilled. First, it is expected that traffic that directly affects the reliability of the production process is treated with the highest priority to meet the reliability and latency requirements. At the same time, *isolation* among different traffic types should be guaranteed, especially if different parties are present on the campus. Likewise, mobility should be properly handled, not only within the campus, but also when moving from the private to the public network. Finally, unexpected events and anomalies should be properly and timely identified, while minimizing the need for human intervention.

Addressing the needs for automated and efficient traffic management is possible by creating virtual logical network instances, i.e., network slices, one for each class of traffic/device that needs to be served on the campus. In this way, each class is assigned to corresponding service level agreements (SLAs) along with isolation and mobility requirements. Appropriate automated mechanisms should be able to translate production requirements into network parameters, such that slices are created automatically and are easy to manage. The solution should also be highly dynamic, for quick commission and decommission of resources when traffic varies, and efficient, so that all traffic can be handled on the same common physical infrastructure. The network should also proactively recognize changes or anomalies in the traffic as well as network behaviors, so that countermeasures are automatically proposed, and effects can be monitored.

ML plays an important role in achieving full automation of slice creation and management. As a matter of fact, different ML methods can be used to support network slicing [10]. As an example, to enable automated slice creation, proper RL mechanisms could be applied to learn the effect of specific slice configuration parameters on the production requirements. Deep reinforcement learning is also a known candidate for efficient resource management [11]. In contrast, unsupervised learning is commonly used for anomaly detection tasks, which are critical in identifying unexpected changes in the production or network environment [12]. So far, in the mentioned references, only data from the network context has been used as input for the ML methods. Additional benefits can be achieved when enriching these approaches with data and requirements from both network and production worlds.

*D. Mobility of Robots and Vehicles*

Flexibility and dynamicity in factories are tightly coupled to the *mobility* of production stations, robots, components and finished goods. AGVs are crucial for flexible factory intralogistics. Primary concerns in this context are *safety* and *security*, i.e., reducing the possibility of accidents and avoiding outages that lead to a disruption of the AGV traffic. Additionally, as they are battery-operated devices, energy-efficiency is an important design criterion. To achieve a high level of autonomy, there is a need for an accurate knowledge of the physical environment, e.g., AGV's own location, nearby people, and (possibly moving) obstacles. With the introduction of 5G systems in industrial facilities, accurate tracking of the radio environment and corresponding QoS in a factory becomes also highly desirable. This is challenging, because factory floors are often characterized by many potentially moving radio frequency (RF) reflectors and opaque RF shields. Although low-power base stations are used, radiated RF energy often remains trapped in a factory hall due to the highly reflective environment, thus creating strong multi-path effects and interference. Consequently, RF conditions can change considerably within a few centimeters [13].

Traditional communication approaches focus on maintaining QoS primarily with reactive methods, e.g., by embedding pilot signals in the transmissions and conducting measurements on the receiver side. However, thanks to recent advances in data-driven ML techniques, there are now tools that can help enhancing future communication systems with proactivity [14]. This applies particularly well in a factory environment, because AGVs and robots have designated pathways and parking areas, move between a predefined number of stations, and have repetitive mobility patterns. ML algorithms, namely correlation analysis and pattern recognition algorithms, can help to find such correlations and patterns in an automated way. Once the future location of a device is known and the radio conditions at that point in space and time can be predicted with a certain accuracy, the communication and network management systems can take advantage of this information. Benefits are expected in terms of enhanced latency and reliability of the communication link, as well as increased number of supported network devices.

The increased number of mobile production machines and AGVs transporting goods also needs to cope with the presence of large obstacles, which potentially require a periodic relocation in the factory. Since such relocations are not random, but follow patterns well-defined by the production processes, ML algorithms can proactively use this knowledge to mitigate interference and link failures, on the network side, as well as optimize the paths of AGVs to avoid unfavorable radio conditions, on the production side.

Finally, ML and especially deep learning algorithms can perform well only if a vast amount of data is available for training on a continuous basis. In this respect, mobile robots and vehicles can be a valuable data source. However, it would defeat the purpose if transmitting this data to the edge cloud (i.e., where it is processed) would compromise the radio resources of the campus network dedicated to production applications. Distributed or federated ML can be exploited to balance the need for computing capabilities close to the data sources and the required bandwidth to transmit this data to a central location. This leads to an additional degree of complexity, e.g., due to the mobility of devices hence changing of the network topology, and at the same time interesting new optimization problems [15].

## III. THE DATA CHALLENGE

One of the driving factors behind the success of machine learning has been the ability of high-capacity learning methods to learn generalizable models from large amounts of data. The success of ML stands and falls on the availability and quality of datasets coming from the specific application domain. Future 5G systems follow a distributed multi-tier design concept in all layers to provide better load management, speed and separation of compute and networking resources. At each tier, large amounts of critical operational and user data are generated, which constitute the so-called *data plane* of 5G systems. Exploiting the full potential from this data requires appropriate strategies to be put in place and design choices to be made to support data collection, data storage, data transfer, data wrangling, ML model training and testing, and lastly implementing those models in the production environment. A current challenge that remains to be solved is the integration of analytics into the network, which is currently complicated due to the various non-standardized interfaces and inconsistent data collection techniques. Initial steps towards enabling such functionalities in 5G systems has been addressed by the network data analytics function (NWDAF) defined in 3GPP TS 29.520. By enabling standard interfaces from the service-based architecture (SBA) to collect data on a subscription/request basis, analytics functionalities can be delivered across the network to enable automation and systematic reporting, thereby addressing the challenges related to custom interfaces.

Campus networks are in general deployed in environments characterized by highly correlated processes that are then reflected in the corresponding manufacturing related data. Augmenting campus network data with such production data has the potential to be greatly beneficial for the ML methods by reducing training time, increasing data efficiency and accuracy, and introducing new insights useful to the learning process. Typical manufacturing related data sources can be provided by production plans, e.g., list of ongoing tasks and their status, goods and production stations from enterprise resource planning (ERP) systems, production planning tools, product lifecycle management software and fleet management tools. Furthermore, location information of end devices, such as AGVs and mobile objects, can be obtained via dedicated localization systems, e.g., based on ultra-wide band (UWB) technology, via lidar and/or video (SLAM) or directly from the 5G system.

The consolidation of manufacturing data with campus network data into a joint ML pipeline poses several challenges regarding data acquisition and integration. Problems arising from, e.g., different time synchronizations/granularities or heterogeneous data types/formats must be addressed to fully leverage potential synergies. Furthermore, access to *real world* data may be difficult in some cases due to, e.g., regulatory policies or the need for costly intrusions into the production processes. Data acquisition in such cases has to be planned in advance and motivated by corresponding ML benefits. In this regard, network simulators/emulators and digital twin representations can serve as synthetic data sources to provide initial performance assessments and enable fast prototyping of ML solutions. Plausibility of synthesized data shall be ensured by calibrating the employed simulators/ emulators with accurate radio propagation models and traffic models, backed, when possible, with real world data and expert knowledge for system parametrization.

## IV. Conclusion

Two emerging technologies, 5G and ML, are essential in supporting vertical industries and facilitating industrial automation. This paper highlights the synergies of these two technologies and details their crucial roles in solving todays major industrial challenges. Those are presented in the form of four key use cases, derived by the overall vision discussed within the KICK project. For each use case, i.e., wireless connectivity and edge-cloud integration, flexibility in reconfiguring the campus network, dynamicity of heterogeneous network services, mobility of robots and vehicles, the paper details the industrial problem, describes the recommended network solution, and justifies the need for ML techniques. Finally, the paper provides valuable insights into current challenges of data acquisition in real systems.

**Ilaria Malanchini** received her Ph.D. degree in electrical engineering from Drexel University, Philadelphia, and Politecnico di Milano in 2011. Currently, she is a senior research engineer at Nokia Bell Labs. Her research interests focus on optimization models, game theory, machine learning, and the application of these techniques to wireless network problems.

**Patrick Agostini** (M.Sc.) currently holds a position as research assistant at Fraunhofer HHI. His main research interests include approximate inference and machine learning with application to wireless communication.

**Khurshid Alam** is a Researcher for Intelligent Networks Department at German Research Center for Artificial Intelligence (DFKI). He received the M. Sc. Degree in Computer Science from the University of Kaiserslautern in 2018. His research interest includes industrial wireless communication, 5G, and software defined networking.

**Michael Baumgart** received his M.Sc. in computer science from the Julius-Maximilians-Universität Würzburg, Germany. In 2020 he joined Infosim as a Consultant R&D. His current research interests are focused on machine learning.

**Martin Kasparick** received the Dipl.-Ing. and the Dr.-Ing. degrees from the Technische Universität Berlin, Germany in 2009 and 2015, respectively. Currently, he leads the Signal and Information Processing Group at Fraunhofer HHI. His research interests include resource allocation, network optimization, and the application of machine learning in wireless communication systems.

**Qi Liao** received her Dr.-Ing degree from Heinrich-Hertz Chair for Information Theory and Theoretical Information Technology, Technische Universität Berlin in 2016. Since 2015, she has been a Senior Research Engineer at Nokia Bell Labs. Her research interests include multi-agent optimization, stochastic process optimization, and machine learning techniques.

**Fabian Lipp** is a Senior Consultant R&D at Infosim since 2018. He received a M.Sc. in computer science from Julius-Maximilians-Universität Würzburg, Germany, in 2014 and after that worked as a research assistant. His current research interest is focused on network management.

**Nikolaj Marchenko** received his Ph.D. degree in Information Technology from University of Klagenfurt, Austria in 2013. Since 2014 he works as Research Engineer at Robert Bosch GmbH. His main topics of research are on industrial IoT and wireless networking for industrial automation.

**Nicola Michailow** received the Dr.-Ing. degree in electrical engineering from Technische Universität Dresden in 2015. From 2015 to 2018, he was with National Instruments. In 2018 he joined Siemens Corporate Technology. His research interests include industrial 5G, software-defined radio and machine learning in wireless. He participated in the projects 5GNOW, CREW, ORCA, IC4F and KICK.

**Rastin Pries** is a research project manager at Nokia. He received his doctoral degree in computer science from the University of Würzburg, Germany in 2010. His main research interests are on applying edge computing to 5G mobile networks as well as on localization approaches for industrial environments.

**Hans Dieter Schotten** is Full Professor and Director of the Institute for Wireless Communications and Navigation at the University of Kaiserslautern and Head of Intelligent Networks department at German Research Center for Artificial Intelligence. He has authored more than 200 papers, filed 13 patents and participated in over 30 European and national projects.

**Slawomir Stanczak** is a Full Professor for network information theory with TU Berlin and the head of the Wireless Communications and Networks department at Fraunhofer Heinrich Hertz Institute (HHI). He is a co-author of two books and more than 200 peer-reviewed journal articles and conference papers in the area of information theory, wireless communications, signal processing and machine learning.

**Stanisław Strzyż** received his M.Sc. in Electronics and Telecommunications from Poznan University of Technology. He has held positions with Nokia and DataX. In 2019 he has joined atesio. He is the co-author of 9 patent applications in the area of heterogeneous networks radio resource management.